\documentstyle[preprint,revtex]{aps}
\begin{document}
\draft
\begin{title} {\Large
Theory of Nuclear Spin-Lattice Relaxation \\
in La$_2$CuO$_4$ at High Temperatures}
\end{title}
\author{{\large A.\ Sokol$^{(1,2)}$, E.\ Gagliano$^{(1)}$,
and S.\ Bacci$^{(1,3)}$}}
\begin{instit} {\normalsize
$^{(1)}$Department of Physics and Materials
Research Laboratory, \\
University of Illinois at Urbana-Champaign,
Urbana, IL 61801-3080 \\
$^{(2)}$L.D. Landau Institute
for Theoretical Physics, Moscow, Russia \\
$^{(3)}$National Center for Supercomputing Applications, \\
University of Illinois at Urbana-Champaign, Urbana, IL 61801-2300
}
\end{instit}
\begin{abstract}
The problem of the nuclear spin-lattice relaxation in
La$_2$CuO$_4$ is revisited in connection with the recent
measurements of the NQR relaxation rate for temperatures up to
$ 900\mbox{K} $ [T.\ Imai {\em et al.},
Phys.\ Rev.\ Lett., in press].
We use an approach based
on the exact diagonalization for the Heisenberg model
to calculate the
short wavelength contribution to the relaxation
rate in the high temperature region, $ T \! \agt \! J/2 $.
It is shown that the spin diffusion accounts for approximately
10\% of the total relaxation rate at $900 \mbox{K} $
and would become dominant for $T\! >\! J$.
The calculated $1/T_1$ is
in good agreement with the experiment both in terms of
the absolute value and temperature dependence.
\end{abstract}
\narrowtext

The fact that the spin dynamics of
the parent insulating compound La$_2$CuO$_4$ is
described by the $S\! =\! 1/2$ Heisenberg model
with $ J \! \simeq \! 1500\mbox{K} $ is now very well
established (for reviews,
see \cite{Chakravarty:review,Manousakis:review}).
Recently, T.\ Imai {\em et al.} \cite{Imai}
have measured the copper nuclear spin-lattice
relaxation rate, $1/T_1$ in
the undoped and Sr-doped La$_2$CuO$_4$ for temperatures
up to $900\mbox{K}$.
They find a plateau in $1/T_1$ as a function of temperature for
$ 700 \! < \! T \! < \! 900\mbox{K} $.
In this temperature region, the
relaxation rate is insensitive to doping, a result which suggests
that at high temperatures
the dominant relaxation mechanism is the same in both metallic
and insulating samples \cite{Imai}.

As it is known, for localized spins,
the relaxation rate is determined by the so-called
``exchange narrowing'' mechanism \cite{Moriya:Anderson}.
The ``exchange narrowing'' here refers to the relaxation process
governed by the spin-spin exchange interaction.
An approach based on the Gaussian
approximation for the dynamic structure factor
has been developed in Ref.\cite{Moriya:Anderson} in order
to calculate the relaxation rate for $ T \! \gg \! J$.
In Ref.\cite{Singh:Gelfand,Gelfand:Singh},
this approach has been combined with high temperature expansion
method and thus extended to finite temperatures of the order of $J$.
For temperatures larger than $J$,
$ 1/T_1 $ has been shown to increase
as the temperature increases.
On the other hand, in the low temperature limit the dominant
contribution to the copper relaxation rate
is due to critical fluctuations around $q\! = \! (\pi/a,\pi/a) $,
and it increases exponentially as the temperature decreases,
$ 1/T_1 \propto T^{3/2}
\exp (2\pi\rho_s / T )$ \cite{Chakravarty:Orbach}.
For $T\! \ll \! J $,
the spin stiffness is $ \rho_s \simeq 0.18 J $.
The interpolation from low to high temperatures shows that
$ 1/T_1 $ as a function of temperature has a minimum.
In Ref.\cite{Chakravarty:Orbach},
its position has been predicted at $ T \simeq 700K $,
a result which seems to be in contradiction with the experimental
data of Ref.\cite{Imai}.

Therefore, the purpose of this work is to understand
whether or not this experimental
result can be {\em quantitatively} understood in the
framework of the nearest-neighbor Heisenberg model.
The analysis of the NMR data in
La$_{2-x}$Sr$_x$CuO$_4$ has lead to the conclusion
that the hyperfine constants in this material approximately
coincide with those of YBa$_2$Cu$_3$O$_x$ \cite{Imai}.
We take advantage of this and use the values of the hyperfine
couplings obtained in Ref.\cite{MPT} for the yttrium-based compounds.
Along with the use of $ J\! \simeq \! 1500\mbox{K} $
for the exchange constant,
this eliminates all adjustable parameters in our calculation.

The copper spin-lattice relaxation rate measured
in the NQR experiment is:
\begin{equation}
\frac{1}{T_1} = \frac{2T}{g^2 \mu_B^2} \, \lim_{\omega \to 0} \,
\frac{\chi_{hf}''(\omega)}{\omega},
\label{T1}
\end{equation}
where
\begin{equation}
\chi_{hf}''(\omega) = \int \frac{d^2{\bf q}}{(2\pi/a)^2}
A^2({\bf q}) \, \chi''({\bf q},\omega)
\label{chi_hf}
\end{equation}
(for simplicity, we use the units where $k_B\! =\! \hbar\! =\! 1$).
In the NQR experiment,
the hyperfine formfactor $A({\bf q})$ is given by \cite{Mila:Rice}:
\begin{equation}
A({\bf q}) = A_{xy} + 2B\cos (q_xa) + 2B \cos (q_ya),
\label{formfactor}
\end{equation}
where $A_{xy} $ and $ B $ are the
in-plane local and isotropic transferred
hyperfine couplings, respectively.
In what follows, we use
$ A_{xy}/B = 0.84 $, $ B = 40.8 \, \mbox{KOe}/\mu_B $ \cite{MPT}.

The relaxation rate of the Heisenberg antiferromagnet has
been discussed in several publications.
However, the low temperature calculation
based on the dynamical scaling theory
\cite{Chakravarty:Orbach} is not valid
for $T\! \sim \! J/2=750\mbox{K}$, where also
the contribution from wave vectors other than
$ q\! =\! (\pi/a,\pi/a) $ becomes important.
On the other hand, it has been mentioned in Ref.\cite{Singh:Gelfand}
that the high temperature expansion
results based on the Gaussian approximation
do not show the low temperature increase of $1/T_1$,
apparently because of the particular functional form
assumed in this calculation for the dynamical
structure factor.
Using large $N$ expansion technique,
it has been shown in Ref.\cite{Chubukov:Sachdev} that
$1/T_1$ is nearly temperature independent
for $T\! \sim \! J/2$.
Unlike our calculation, this approach does not start from the
$ S\! =\! 1/2$ lattice model and so the absolute value of $1/T_1$
is evaluated in Ref.\cite{Chubukov:Sachdev}
using the low-temperature fit
of the {\em same} data
and not the hyperfine couplings $A_{xy},B$.

For $T\! \ll \! J$,
the spin diffusion ($ q \! \to \! 0 $) contribution to the relaxation
rate, $ (1/T_1)_{diff} $,
is negligible because the spin diffusion constant, $D$,
is exponentially
large \cite{Chakravarty:Orbach,Chakravarty:Gelfand:etc}.
However, $D$ rapidly decreases as the temperature increases,
that is, the $ q \! \to \! 0 $ component may be important
for higher temperatures.
In a pure two-dimensional model,
the conservation of spin leads to the
divergence of $ (1/T_1)_{diff} $;
that is, the relaxation would be faster than exponential.
However, in a real system $ (1/T_1)_{diff} $
remains finite and its magnitude
is determined by the length scale $ L_s $,
set either by spin-nonconserving forces or
three-dimensional effects.
Since in any cluster calculation
(exact diagonalization or Monte-Carlo) the cutoff is set
by the lattice size, we have taken
into account the $q\! \to \! 0$ contribution separately.

Our approach for the calculation of the
short wavelength contribution to the relaxation rate
is based on the exact diagonalization of the Hamiltonian
for the $ 4 \! \times \! 4 $ cluster.
Since the nuclear spin-lattice relaxation rate is determined by
short-range spin correlations, our results
are relevant to the
real system as long as the correlation length is not large compared
to the cluster size.
The spectral representation for $ \chi_{hf}'' $
can be written in terms of the Hamiltonian eigensystem as follows:
\begin{eqnarray}
\frac{\chi_{hf}''(\omega )}{g^2 \mu_B^2} & = & \frac{\pi}{Z} \,
\sum_{ab}
\left[ \exp(-E_a/T) - \exp(-E_b/T) \right]
\nonumber \\
& \times &
\delta ( E_a \! - \! E_b \! + \! \omega ) \,
\frac{1}{N} \sum_{q\neq 0}
A^2({\bf q}) \, | \langle a | S^z_q | b \rangle |^2,
\label{spectral}
\end{eqnarray}
where $ E_{a,b} $ are the eigenvalues of the Hamiltonian and
$ Z = \sum_a \exp (-E_a/T) $ is the partition function.
In the thermodynamic limit ($N\! \to \! \infty $),
$ \chi_{hf}'' $ is a continuous function of frequency, while
for finite size it
is a superposition of delta functions.

For a finite cluster, the limit $\omega \! \to \! 0 $
in Eq.(\ref{T1}) is not defined,
but we argue that the thermodynamic $ \chi''_{hf}(\omega) $
can be calculated using the following procedure.
Consider the auxiliary function $ I_N(\omega) $ given, for a cluster
of size $N$, by
\begin{equation}
g^2 \mu_B^2 I_N(\omega) = \frac{1}{2} \, \int_{-\omega}^\omega
d\epsilon \ \frac{ \chi_{hf}''(\epsilon)}{\epsilon}.
\end{equation}
{}From this equation,
$ \chi_{hf}''(\omega)/\omega = g^2 \mu_B^2 (d I_N /d\omega) $.
For finite cluster, $ I_N(\omega) $
can easily be calculated from
the eigenstates of the Hamiltonian:
\begin{equation}
I_N(\omega ) =
\sum_{ab} I_{ab}
\left[ \theta \! \left( E_a \! - \! E_b \! + \! \omega \right)
\! - \!
\theta \! \left( E_a \! - \! E_b \! - \! \omega
\right) \right],
\end{equation}
where $\theta(x) $ is the Heaviside function and
\begin{eqnarray}
I_{ab} & = & \frac{\pi}{2Z}
\frac{\exp(-E_a/T) - \exp(-E_b/T)}{E_b-E_a}
\nonumber \\
& \times & \frac{1}{N} \sum_{q\neq 0} A^2({\bf q})
| \langle a | S^z_q | b \rangle |^2.
\end{eqnarray}
(for $a\! =\! b$ we take the limit $E_a\! \to \! E_b$).
The auxiliary function $ I_N(\omega) $
is quite smooth as long as the temperature is not much smaller than
the gap between the ground state and the rest of the spectrum,
which for the 16-site cluster is of order $J/2$.
For temperatures $ T \! > \! 1.5\! - \! 2 \, J $,
we find no appreciable size dependence:
$ I_{10} \! \simeq \! I_{16} $.
In the study of static properties of the Heisenberg model
\cite{Bacci:Gagliano:Dagotto}, no discrepancy was found
between the $4\! \times \! 4$ cluster and Monte-Carlo
results for larger systems at $ T\! > \! J $.
Both the discrepancy and the errorbars in the fitting
of $I_{16}$ by
a smooth function increase up to approximately 10\% for
$T\! \simeq \! J/2 $. Thus, we will
assume that our calculation of the short wavelength contribution
to $1/T_1$ has 10\% accuracy.

Now we turn to the calculation of the $q \! \to \! 0$
contribution to $1/T_1$. For $ L_s^{-1} < q < \max(\xi,a)^{-1} $
and $ \omega\tau \ll 1 $,
the dynamical spin
susceptibility $ \chi({\bf q},\omega) $ has the following form:
\begin{equation}
\chi({\bf q}, \omega ) = \chi ({\bf q}) \,
\frac{D q^2}{D q^2 - i \omega},
\end{equation}
where $D$ is the diffusion constant and  $ \tau $
a characteristic relaxation time.
Substituting this expression into Eq.(\ref{T1}), we obtain
\begin{equation}
\left( \frac{1}{T_1} \right)_{diff} =
\frac{T \chi_0 a^2 A^2(q\! =\! 0)}{\pi g^2 \mu_B^2 D} \,
\log \frac{L_s}{L_{f.s.}},
\label{T1diff}
\end{equation}
where we take $ L_{f.s.} \! > \! \xi  $
to be equal to the size of our cluster.
For $ T \gg J $, the diffusion constant is
$ D \simeq 0.43 Ja^2 $ \cite{Morita},
so that $ (1/T_1)_{diff} \sim
7400 \times \log(L_s/L_{f.s.}) \ \mbox{sec}^{-1} $ is
at least several times larger
than the measured rate at the maximal accessible temperature
$ 900\mbox{K} $ \cite{Imai}. This contribution is larger than
the calculated short wavelength contribution
at the same temperature.
Therefore, the relaxation rate of the 2D Heisenberg model
for $ T \! > \! J $
is a {\em poorly defined quantity}, since
it strongly depends on the way the cutoff
is taken into account.
In this temperature region,
an accurate calculation of $1/T_1$ would have
to involve the actual mechanism destroying the diffusion.
However, since temperatures larger than
the exchange constant are not experimentally
accessible in La$_2$CuO$_4$,
we will examine now whether or not the spin diffusion
substantially contributes to the relaxation rate
at $600\! -\! 900\mbox{K}$.

In order to address this issue, we have to determine
both the diffusion
constant $D$ and the length scale $L_s$ for $T \! \simeq \! J/2$.
The diffusion constant is estimated
as \cite{deGennes,Bennett:Martin}:
\begin{equation}
D = \gamma \lim_{q\to 0} q^{-2} \langle
\omega^{2}_\chi \rangle^{3/2}_q
\langle \omega^{4}_\chi \rangle^{-1/2}_q,
\label{Dest}
\end{equation}
where $\langle \omega^{2n}_\chi \rangle_{q} $ are
the frequency moments of the dynamical response function,
\begin{equation}
\langle \omega^{2n}_\chi \rangle_{q} =
  \frac{\int \omega^{2n-1} \chi''({\bf q},\omega) \, d\omega }{
    \int \omega^{-1} \chi''({\bf q},\omega) \, d\omega },
\end{equation}
and $ \gamma $ is a numerical factor which depends
on the assumed short time relaxational
behavior \cite{Reiter}.
Taking $ \gamma = \sqrt{\pi/2} \approx 1.25 $
\cite{Bennett:Martin}
yields $ D_{T=\infty} = 0.40 Ja^2 $, which is quite close to the
value $ 0.43 Ja^2 $ \cite{Morita} obtained through an evaluation
of the memory function.
In Ref.\cite{T-K:McF:Collins}, general expressions for the series
in $\beta\! =\! J/T$ for $ \langle \omega^{2n}_\chi \rangle_{q} $
have been derived. Using these results, we calculate
first two terms of the high temperature
expansion for the diffusion constant:
\begin{equation}
\frac{D}{Ja^2} = \frac{\sqrt{\pi}}{2\sqrt{5}} +
\frac{21 \sqrt{\pi} \beta}{40\sqrt{5}} + {\rm O}(\beta^2)
\approx 0.40 + 0.42 \beta + {\rm O}(\beta^2).
\end{equation}
Two leading terms in the high temperature expansion series are
not sufficient for the accurate estimate of $D$ at
$T\! \sim \! J/2 $. However, we know that the diffusion constant
should scale approximately as $ D \propto \xi $ in the quantum
critical region, $ \rho_s \! < \! T \! < \! J $. Using
Monte-Carlo data of Ref.\cite{Ding:Makivic}
for the correlation length,
$ \xi(T) $, we estimate $ D \sim 3 Ja^2 $
at $ T\! = \! 900\mbox{K}$, which when
substituted into Eq.(\ref{T1diff}) gives
$ (1/T_1)_{diff} \sim (200\! -\! 300) \times
\log(L_s/L_{f.s.}) \ \mbox{sec}^{-1} $.

Now we turn to the evaluation of the logarithm
in Eq.(\ref{T1diff}). Since the hyperfine splitting
$ \sim 1.5 \cdot 10^{-7} \mbox{eV} $  is very small and
above the tetragonal-to-orthorhombic transition temperature,
$ T_{T-O} \simeq 525\mbox{K} $, the Dzyaloshinskii-Moriya
interaction vanishes, the cutoff is determined either by the
three dimensional effects or by the nonconservation of spin.
Consider first
the cutoff due to the three-dimensional effects, $ L_s^{3D} $,
which is set by the
interplanar diffusion constant, $ D_\perp $.
For estimation purposes, we express
$D$ in terms of the characteristic damping of spin waves
for small wave vectors, $ \bar{\gamma} $ \cite{Halperin:Hohenberg}.
With omission of all factors of the order of unity,
we get $ D_\perp/D_\parallel \sim J'/J $,
which yields $ L_s^{3D} \sim 300a $, a quite large value.

Given the size of $L_s^{3D} $, we consider an alternative
physical origin for the cutoff,
the presence of weak disorder in CuO$_2$ planes.
For temperatures above $700\mbox{K}$,
the oxygen content changes after the heating cycle
by approximately $ 0.004 $ per unit cell \cite{Imai};
that is, the average distance
between nonstoechiometric oxygen atoms,
which we identify with $L_s$,
is $10\! -\! 20a$.
Although the value of $L_s$ cannot
be determined quite accurately,
the $q\! \to \! 0 $ contribution to $1/T_1$
depends on $L_s$ only weakly. In what follows,
we plot the results for $L_s\! =\! 10a$ and $L_s\! =\! 20a$.
Substituting the above values of $L_s$ into Eq.(\ref{T1diff}),
one obtains
that the spin diffusion contribution accounts for approximately 10\%
of the total spin lattice
relaxation rate for $T\! =\! 900\mbox{K}$, but rapidly
decreases as the temperature decreases.
This explains why
the tetragonal-to-orthorhombic transition at $T_{T-O}=525K$
does not have any observable effect on the spin lattice relaxation
although it affects $L_s$.

The total relaxation rate for $L_s\! =\! 10$ and $L_s\! =\! 20$
and the short wavelength contribution
alone are plotted in Fig.\ref{Fig1}
together with the experimental result of Ref.\cite{Imai}.
The theoretical result is in 15\% agreement with the experiment
(for $L_s\! =\! 10$). The agreement can be improved by either taking
smaller $L_s$, or changing the hyperfine constants by 7\%
(actually, $A_{xy},B$ are known only with 5-10\%
accuracy \cite{MPT}).
It is important to emphasize, however, that the ambiguity
in definition
of $1/T_1$ as a function of the cutoff $L_s$ exceeds our estimate
of the systematic
error of the finite cluster calculation; moreover, the cutoff itself
can not be determined quite accurately.
The spin diffusion ($q\! \to \! 0 $)
contribution rapidly increases as the temperature increases
(Fig.\ref{Fig1}, inset) and becomes dominant for $T \! > \! 1.5J$,
as it is shown on Fig.\ref{Fig2}.
Although this temperature range
is beyond the limit of chemical
stability for La$_2$CuO$_4$, it may be of interest for other
materials
described by the Heisenberg model but with smaller $J$, such as
Cu(HCO$_2$)$_2 \cdot $4H$_2$O
and Cu(pyz)$_2$(ClO$_4$)$_2$ \cite{lowJ}.

To summarize, we have calculated the copper spin-lattice relaxation
rate for La$_2$CuO$_4$ without introducing
any adjustable parameters.
The spin diffusion ($q\! \to \! 0 $) contribution
is shown to account for
10\% of the relaxation for the maximal temperature achieved in
the experiment, $900\mbox{K}$, although it would become
dominant for larger
temperatures, thereby explaining the discrepancy between
different calculations of the relaxation rate at high temperatures.
The measured $1/T_1$ \cite{Imai} turned out to be quantitatively
consistent with the nearest-neighbor Heisenberg model
description of the
spin dynamics in La$_2$CuO$_4$.

Thanks are due to S.J.\ Clarke, D.\ Frenkel, M.P.\ Gelfand,
L.P.\ Gor'kov, C.P.\ Landee, D.\ Pines, G.\ Reiter,
and C.P.\ Slichter for many stimulating discussions,
to M.P.\ Gelfand and R.R.P.\ Singh for communicating
their results before publication, and to D.\ Pines for suggestions
about the manuscript.
One of the authors (A.S.) is indebted to A.V.\ Chubukov for
numerous conversations on the low dimensional critical phenomena,
and to T.\ Imai for many discussions on the experimental
aspects of the nuclear magnetic resonance.
This work has been supported
by the NSF Grant DMR89-20538 through the Materials
Research Laboratory. The computer calculations were
performed on the Cray Y-MP at the National Center for
Supercomputing Applications, Urbana, Illinois.

\figure{
The calculated $1/T_1$ as a function of temperature
without (solid line) and with the spin diffusion contribution for
$L_s=10$ (dashed line), and for $L_s=20$ (dotted line).
Dots are the experimental result of Ref.\cite{Imai}.
The errorbars due to the calculation inaccuracy
(less than 10\%, not shown) are smaller than
the ambiguity in the definition of $1/T_1$
related to $q\! \to \! 0 $ cutoff.
Inset: the spin diffusion ($q\! \to \! 0 $) contribution
to the relaxation rate.
\label{Fig1}}

\figure{
Same as Fig.\ref{Fig1}, but in the temperature range
$ J/2 \! < \! T \! < \! 3J $ for a hypothetical
heat-resistant sample.
\label{Fig2}}


\begin{references}

\bibitem{Chakravarty:review}
S. Chakravarty, in Proceedings of
{\em High Temperature Superconductivity}, edited by K.S. Bedell
{\em et al.} (Addison-Wesley, CA, 1990).

\bibitem{Manousakis:review}
E. Manousakis, Rev. Mod. Phys. {\bf 63}, 1 (1991).

\bibitem{Imai}
T. Imai, C.P. Slichter, K. Yoshimura, and K. Kosuge,
Phys. Rev. Lett., in press.

\bibitem{Moriya:Anderson}
T. Moriya, Progr. of Theor. Phys. {\bf 16}, 641 (1956);
see also P.W. Anderson, Rev. Mod. Phys. {\bf 25}, 269 (1953).

\bibitem{Singh:Gelfand}
R.R.P. Singh and M.P. Gelfand, Phys. Rev. B, {\bf 42}, 996 (1990).

\bibitem{Gelfand:Singh}
M.P. Gelfand and R.R.P. Singh, preprint.

\bibitem{Chakravarty:Orbach}
S. Chakravarty and R. Orbach, Phys. Rev. Lett.,
{\bf 64}, 224 (1990).

\bibitem{MPT}
H. Monien, D. Pines, and M. Takigawa,
Phys. Rev. B {\bf 43}, 258 (1991).

\bibitem{Mila:Rice}
F. Mila and T.M. Rice,
Physica C, {\bf 157}, 561 (1989).

\bibitem{Chubukov:Sachdev}
A.V. Chubukov and S. Sachdev, preprint.

\bibitem{Chakravarty:Gelfand:etc}
S. Chakravarty, M.P. Gelfand, P. Kopietz, R. Orbach, and
M. Wollensak, Phys. Rev. B {\bf 43}, 2796 (1991).

\bibitem{Bacci:Gagliano:Dagotto}
S. Bacci, E. Gagliano, and E. Dagotto, Phys. Rev. B
{\bf 44} 285 (1991).

\bibitem{Morita}
T. Morita, Phys. Rev. B {\bf 6}, 3385 (1972).

\bibitem{deGennes}
P.G. De Gennes, J. Phys. Chem. Solids, {\bf 4}, 223 (1958).

\bibitem{Bennett:Martin}
H.S. Bennett and P.C. Martin, Phys. Rev. {\bf 138}, A608 (1965).

\bibitem{Reiter}
Note, that the calculation for the classical model [O.F. de Alcantara
Bonfim and G. Reiter, preprint] would not be applicable
to $S\! = \! 1/2$ problem for $T\! \agt \! J/2$.

\bibitem{T-K:McF:Collins}
R.A. Tahir-Kheli and D.G. McFadden, Phys. Rev. B.
{\bf 1}, 3178 (1970);
M.F. Collins, Phys. Rev. B {\bf 4}, 1588 (1971).

\bibitem{Ding:Makivic}
H.-Q. Ding and M.S. Makivic, Phys. Rev. Lett. {\bf 64}, 1449 (1990).

\bibitem{Halperin:Hohenberg}
B.I. Halperin and P.C. Hohenberg, Phys. Rev. {\bf 188}, 898 (1969);
A.V. Chubukov, Phys. Rev. B {\bf 44}, 12318 (1991).

\bibitem{lowJ}
S.J. Clarke and C.P. Landee, private communications.

\end{references}
\end{document}